\documentclass[12pt]{emulateapj}
\usepackage{amsmath}
\usepackage{txfonts}
\usepackage{graphicx}
\usepackage{dcolumn}
\usepackage{bm}
\usepackage{flafter}
\usepackage{mathrsfs}

 \shorttitle{Planetesimal dynamics in inclined binary systems}
\shortauthors{ZHAO ET AL.}

\begin{document}

\title{Planetesimal dynamics in inclined binary systems: the role of gas-disk gravity}

\author{Gang Zhao\altaffilmark{1}, Ji-Wei Xie\altaffilmark{$^{\star}$1, 2}, Ji-Lin Zhou\altaffilmark{1} and Douglas N.C. Lin\altaffilmark{3, 4}}
\affil{$^1$Department of Astronomy \& Key Laboratory of Modern
Astronomy and Astrophysics in Ministry of Education, Nanjing
University, Nanjing, China 210093.} \affil{$^2$Department of
Astronomy and Astrophysics, University of Toronto, Toronto, ON M5S
3H4, Canada;} \affil{$^3$UCO/Lick Observatory, University of
California, Santa Cruz, CA 95064} \affil{$^4$Kavli Institute for
Astronomy and Astrophysics, Peking University, Beijing, China}
\altaffiltext{$\star$}{Corresponding to xiejiwei@gmail.com}

\begin{abstract}
We investigate the effects of gas-disk gravity on the planetesimal
dynamics in inclined binary systems, where the circumprimary disk
plane is tilted by a significant angle ($i_B$) with respect to the
binary disk plane. Our focus is on the Lidov-Kozai mechanism and the
evolution of planetesimal eccentricity and inclination. Using both
analytical and numerical methods, we find that, on one hand, the
disk gravity generally narrows down the Kozai-on region, i.e., the
Lidov-Kozai effect can be suppressed in certain parts of (or even
the whole of) the disk, depending on various parameters. In the
Kozai-off region, planetesimals would move on orbits close to the
mid-plane of gas-disk, with the relative angle ($i^{'}$) following a
small amplitude periodical oscillation. On the other hand, when we
include the effects of disk gravity, we find that the Lidov-Kozai
effect can operate even at arbitrarily low inclinations ($i_B$),
although lower $i_B$ leads to a smaller Kozai-on region.
Furthermore, in the Kozai-on region, most planetesimals'
eccentricities can be excited to extremely high values ($\sim 1$),
and such extreme high eccentricities usually accompany orbital
flipping, i.e., planetesimal orbit flips back and forth between
anterograde and retrograde. Once a planetesimal reaches very high
orbital eccentricity, gas drag damping will shrink the planetesimal
orbit, forming a ``hot planetesimal'' on a near circular orbit very
close to the primary star. Such a mechanism, if replacing the
planetesimals and gas drag damping with Jupiters and tidal damping
respectively, may lead to frequent production of hot-Jupiters..
\end{abstract}
\section{Introduction}
As of today, over 60 exoplanets have been found in binary star
systems, and current observations show that the multiplicity rate of
the detected exoplanet host stars is around 17\% \citep{MN 09,Egg
10}.  Planet formation in binary system systems presents numerous
challenges, as each stage of the planet formation process can be
affected by the binary companion. A crucial stage that may be
particularly sensitive to binary effects is the accumulation of
1-100 km-sized planetesimals (see the review by \citet{Hag 10} and
the references therein). Because of the perturbations from the
binary companion, planetesimals will be excited to orbits with high
relative velocities, preventing or even ceasing their growth
\citep{Hep 78,Whi 98}. In the past decade, with several discoveries
of exoplanets in close binary of separation $\sim$ 20 AU \citep{Que
00, Hat 03,Zuc 04, Cor 08, Cha 11}, the issue of planetesimal growth
in binary systems becomes more challenging and therefore attracts
many researchers as well as many dynamical and collisional studies
\citep{MS 00, MN 04, The 04, The 06, The 08, The 09, The 11,Paa 08, Sch 07, PL 10,
KN 08, Bea 10, Giu 11, XZ 08, XZ 09, Xie 10a, Xie 10b}.

Most of previous studies had considered only coplanar or
near-coplanar cases, where the tilted angle between the binary
orbital plane and the circumprimary disk plane was close to zero,
i.e., $i_B\sim0$.  In fact, the coplanar case is reasonable only if
it is applied to relatively close binary systems with separation
less than $\sim$ 40-200 AU \citep{Hal 94, Jen 04}, beyond which the distribution
of $i_B$ is likely to be random and therefore the highly inclined
case is more relevant. Planetesimal dynamics in highly inclined
binary systems have only been investigated by \citet{Mar 09b}, and
most recently (at the time of writing this paper) by \citet{Xie 11},
\citet{Fra 11}, and \citet{Bat 11}.

\citet{Mar 09b} found that, due to the perturbations of a inclined
binary companion, planetesimals' nodal lines became progressively
randomized, raising their relative velocities to the degree that
planetesimal growth by mutual collision was significantly prevented.
Nevertheless, the gaseous protoplanetary disk was ignored in their
study, where planetesimals were only subject to the gravity of the
binary stars. In reality, the gaseous disk can generally have
crucial effects on planetesimal dynamics through two factors. One is
the hydrodynamic drag force, which has been investigated in detail
by \citet{Xie 11}.  When gas drag is included, it is found that
planetesimals from the outer regions (where conditions are hostile
to planetesimal accretion) jump inward into an accretion-friendly
region and pile-up there. This is referred to as the planetesimal
jumping-piling effect (PJP), and its general result, as shown in
\citet{Xie 11}, is to form a severely truncated and dense
planetesimal disk around the primary, providing conditions which are
favorable for planetesimal growth and potentially allow for the
subsequent formation of planets.  Another crucial factor is the
gravity of gaseous disk, which has been studied recently by
\citet{Fra 11} with a hydro-dynamical model and by \citet{Bat 11}
with an analytical model. Generally, it is found that the gravity
could pull the planetesimals back towards the middle plane of
gas-disk. With proper conditions, such as a massive gas disk and/or
a large binary distance, Lidov-Kozai effect could be suppressed
regardless of $i_{B}$. However, \citet{Fra 11} could only focus on
several typical cases with a few planetesimals in relative short
simulation timescale because of the large computational hours, while
\citet{Bat 11} only concentrated on cases of very wide binaries with
separation of $\sim$ 1000 AU, aiming to just identify the important
physical processes at play.

In this paper, we investigate the effects of gas-disk gravity on
planetesimal dynamics in inclined binary systems through both
analytical and numerical fashions. Analytically, we derived the
condition at which Lidov-Kozai effect is turned off by the disk gravity.
Numerically, we confirm our analytical results and provide a global
quantitative view of the role of gas-disk gravity in a large
parameter space. Furthermore, one specific attention is given to the
role of disk gravity in shaping the PJP effect found by \citet{Xie 11}. The paper is outlined as follows.

In section 2, we describe our disk model and the initial set up. In
section 3, we analyze the secular motion of a planetesimal under the
gravity of disk and stars, focusing on the evolution of planetesimal
inclination and eccentricity. The analytical study is followed by
the numerical simulations presented in section 4. In section 5, we
discuss some issues, including the PJP effect, implication for
hot-Jupiters and disk precession. Then in Section 6 we present our
summary.

\section{Disk Model}
In our model, planetesimals are assumed to be initially moving on
circular orbits in the mid-plane \footnote{This is equivalent to assume that
their proper eccentricities (and inclinations) are equal to the
forced one.} of a gaseous disk around the
central star with one solar mass ($M_{\odot}$). A companion star
with mass of $M_{B}$ (a free parameter) is orbiting around the
central star-disk system with an orbital semimajor axis of $a_{B}$
(a free parameter), inclination of $i_{B}$ (a free parameter,
relative to the mid-plane of the disk), eccentricity of $e_{B}=0$
(constant). In this paper, for simplicity, we only consider the circular case
($e_{B}=0$) and focus on the effect of gas gravity. For the
eccentric case $e_{B}\ne0$, the Lidov-Kozai mechanism itself is more
complicated \citep{Kat 11, LN 11}, and thus this case \footnote{If $e_B=0$, the average Hamiltonian is axisymmetric, thus the vertical angular momentum is an integral of motion, and the planetesimal orbit can be well described with the classic Kozai effect. Otherwise, if $e_B>0$, the vertical angular momentum is not a constant any more, and the classic Kozai effect should be modified with the so called Eccentic Kozai effect \citep{LN 11}. }  is not
addressed in this paper.

For the gas disk, we use a 3-dimension steady model as in \citet{TL
02}. In cylindrical coordinates $(r,z)$, the disk density profile is
\begin{equation}\label{profile}
\rho_g(r,z)=\rho_0f_g\left(\frac{r}{\rm{AU}}\right)^\beta\exp\left(-\frac{z^2}{2h_g^2}\right),
\end{equation}
and the gas rotation rate is
\begin{equation}\label{Omega}
\Omega_g(r,z)=\Omega_\textrm{K,mid}\left[1+\frac{1}{2}\left(\frac{h_g}{r}\right)^2\left(\beta+\gamma+\frac{\gamma}{2}
\frac{z^2}{h_g^2}\right)\right],
\end{equation}
where $\Omega_\textrm{K,mid}$ is the
Keplerian rotation in the mid-plane, $h_g(r)=h_0(r/{\rm{AU}})^{(\gamma+3)/2}$ is the scale height of
gas disk, $f_g$ is a scaling number with respect to the minimum mass
of solar nebulae  (\citet{Hay 81}, MMSN hereafter), $\rho_0=2.83
\times 10^{-10}\rm{gcm^{-3}}$, $\gamma=-0.5$,
$h_0=0.33\times10^{-2}$, and $\beta$ is a free parameter. The surface density of the disk
has a power-low form of
$\Sigma_g=\sqrt{2\pi}\rho_0f_gh_0(r/\rm{AU})^{k}$, where $k=\beta+1.25$. Nominally, in this paper, we set $k=-1$ as the stander case. The inner and outer boundaries of the disk are set as $r_{\rm in}=0.1$ AU and
$r_{\rm out}=12.5$ AU. Their values have little effect on the final
results as long as they are not very close to the planetesimals.

Our disk model is a very simple one, which ignores the reaction of
gas disk to the binary perturbations. In more realistic situations,
as shown in the simulations of  \citet{Lar 96, FN 10}, the disk
will become eccentric, develop a warp and precess under the
perturbations of the companion star. Nevertheless, as pointed out by
\citet{Fra 11} (also see our discussion in section 5.2) ,
planetesimals' secular dynamical behaviors are similar both in the
evolving and non-evolving disk models, and thus our choice of a
steady model can be reasonable to at least a zeroth order
approximation. Furthermore, using such a simple gas disk model is
much less time consuming in computing the disk gravity as compared
to using a hydrodynamical code, allowing us to see the effect of
disk gravity on a much longer timescale.  In addition, our simple
gas disk model is convenient for making some analytical studies.

It also worthy noting that the gaseous disk would slowly
relax to the binary orbital plane on the viscous evolution timescale \citep{FN 10}. Thus the assumption of a constant and relatively large $i_{B}$ in our model is only relevant if the viscous timescale, $t_{vis}\sim r^{2}/(\alpha h^{2} \Omega_{\rm k,mid})$, is larger than the secular perturbation timescale, $t_{sec}\sim2\pi/B$. Equating these two timescales, the critical viscous parameter can be derived as
\begin{equation}\label{profile}
\alpha_{c}\sim5\times10^{-2}\left(\frac{r}{\rm{10 AU}}\right)^{5/2}\left(\frac{a_{B}}{\rm{50 AU}}\right)^{-3}\left(\frac{M_{B}}{\rm{M_{\odot}}}\right)^{1/2}.
\end{equation}
Therefore, a high inclined case, which studied in this paper, is relevant only for $\alpha<\alpha_{c}$. If otherwise, $\alpha>\alpha_{c}$, it is likely to reduce to a near coplanar case, which has been studied in many previous works \citep{MS 00, The 06, Paa 08, XZ 08, XZ 09}.

\section{Analytic study}
In this section, we analytically study the secular dynamics of a
planetesimal under the gravitational perturbations from both the
companion star and the disk.  Our interests focus on the evolution
of the planetesimal's orbital eccentricity and inclination, aiming
to see how the Lidov-Kozai effect operates if the disk gravity is
included. For the sake of this derivation, we introduce two
coordinate systems: (1) the disk coordinate, where the
$X^{'}Y^{'}$-plane is set as the disk mid-plane with the $X^{'}$
direction towards the ascending node of the binary orbit, and (2)
the binary coordinate, where the $XY$-plane is set as the orbital
plane of the binary star with the $X$ direction the same as $X^{'}$.
In the disk coordinate system, angular elements are marked with a
superscript (`` $'$ ''). For example, $i^{'}$ and $\Omega^{'}$
denote the orbital inclination and longitude of ascending node in
the disk coordinate system respectively, while $i$ and $\Omega$ are
those in the binary coordinate system.

\subsection{The Disturbing Function}
The disturbing function of the star-disk-planet system can be
expressed as
\begin{equation}\label{rall}
R=R_D+R_B,
\end{equation}
where $R_D$ and $R_B$ are contributions from the gravity of disk and
binary stars, respectively.

According to \citet{Nag 00} (see the appendix of their paper),
taking the second order approximation, $R_{D}$ can be expressed as
\begin{equation}\label{rd}
R_{D}=-\frac{na^2}{2}\left[Te^2+Si'^2\right],
\end{equation}
where $n$, $a$, $e$ and $i'$ are the orbital mean frequency,
semi-major axis, eccentricity, inclination (in disk coordinate) of
the planetesimal.
$T$ and $S$ are two characteristic frequencies (see the appendix of
this paper for details of their definition and calculation) which,
under the disk model assumed in section 2, can be approximately fit by the following formulas.
\begin{eqnarray}\label{ral2}
T(a)&=&f_g\times 4.5\times10^{-4}\left(\frac{a}{\rm AU}\right)^{k+1} \rm rad/yr,\\
S(a)&=&f_g\times 1.7\times10^{-2}\left(\frac{a}{\rm AU}\right)^{k+1/4} \rm rad/yr.\nonumber
\end{eqnarray}
Note, $T$ is actually the apsidal recession rate of a planetesimal
if the planetesimal is affected only by the disk gravity in the
coplanar case ($i^{'}=0$). Our calculation of $T$ is generally
consistent with that of \citet{Bat 11} (here $f_{g}=1$ corresponds
to a disk mass of $\sim0.02 \rm M_{\odot}$ in the figure 2 of their
paper) who used a similar disk model but different computing
technics. However, we emphasize that $T$ should be scaled with the
local surface density as in equation (\ref{ral2}) rather than with
the total mass of the disk (as was done in figure 2 of \citet{Bat
11} and Eqn. (30) in \citet{Fra 11}).

Following \citet{Inn 97}, the binary part of the disturbing function can be expressed
as:
\begin{equation}\label{rb}
R_{B}=\frac{na^2}{2}B\left[e^2-(1+4e^2-5e^2\cos^2\omega)\sin^2i\right],
\end{equation}
where $i$ and $\omega$ denote the orbital inclination and pericenter
(in binary coordinate) of the planetesimal. The characteristic
frequency $B$ is actually the precession rate of the planetesimal
caused by the secular binary perturbation in the  coplanar case
($i=0$), and in the first order it can be expressed as
\begin{equation}\label{rbcoe}
B\sim B_1=\frac{3GM_B}{4na_B^3(1-e_B^2)^{3/2}}.
\end{equation}
However, such a first order expression can be rather inaccurate
unless one uses the second order correction ($B_2$) as suggested by
\citet{The 06} and \citet{Giu 11},
\begin{equation}\label{rbcoepie}
B\sim B_{2}=B_{1}\left[1+\frac{32M_B}{M(1-e_B^2)^3}\left(\frac{a}{a_B}\right)^2\right].
\end{equation}
Hereafter, we adopt $B=B_{2}$ if there is no specific explanation.

\begin{figure}
\begin{center}
\includegraphics[scale=0.75]{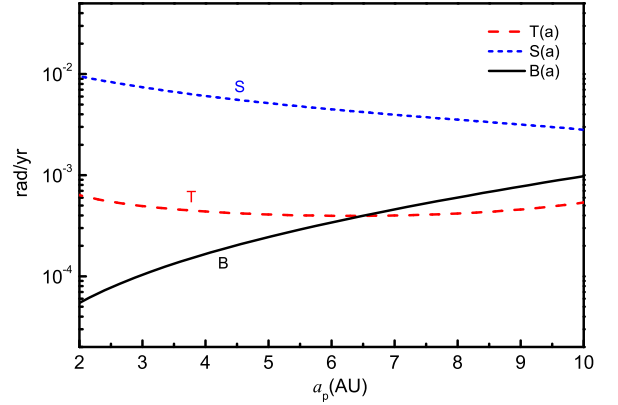}
\caption{The values of $T(a)$, $S(a)$, and $B(a)$. The blue dotted
line and red dashed line indicate $T(a)$ and $S(a)$ of an MMSN disk
with inner edge 0.1AU and outer edge 12.5AU and surface density slope of $k=-1$. The black solid line
shows $B(a)$ of a companion star at $a_{B}=50$ AU with 0.5$M_\odot$.}
\label{tbs}
\end{center}
\end{figure}

We plot $T$, $S$, and $B$ in Figure \ref{tbs} for the standard case,
where the companion has mass of $M_{B}=0.5 M_\odot$, semimajor axis
of $a_{B}= 50$ AU, and the disk surface density slope of $k=-1$. The blue
doted line, red dashed line, and black solid line indicate $T$, $S$
and $B$ as a function of the semimajor axis of the planetesimal
($a$), respectively. As can be seen, $S$ is much greater than $T$
and $B$ in the whole of the plotted region of the disk, while $T$ is
greater (less) than $B$ in the inner (outer) region. We will show in
the following subsections that such a picture of $T$, $B$ and $S$
determines the dynamical evolution of the planetesimal's orbit.

\subsection{Evolution of the Planetesimal Inclination}
As the planetesimal is initially moving on a circular orbit in the
mid-plane of the disk, the initial $e$ and $i^{'}$ are approximately
zero, thus we ignore quantities that are on an order of higher than
$o(e^{2})$, $o(i^{'2})$ or $o(ei^{'})$. The disturbing function
relating to the inclination then can be reduced to
\begin{equation}\label{rishort}
R_{rd}\sim\frac{na^2}{2}\left[-Si'^2-B\sin^2i\right].
\end{equation}
Considering the relation between $i$, $i^{'}$ and $\Omega^{'}$ and
introducing two new variables $p=i'\sin \Omega'$ and $q=i' \cos
\Omega'$, then Lagrange's planetary equations (relating to $i^{'}$
and $\Omega^{'}$) can be written as (see the appendix for detailed
derivation)
\begin{eqnarray}
\label{dpdq}
\frac{dp}{dt}&=&-\left(B\cos2i_B+S\right)q+\frac{B}{2}\sin2i_B,\nonumber\\
\frac{dq}{dt}&=&\left(B\cos^2i_B+S\right)p,
\end{eqnarray}
where $i_{B}$ is the angle between the disk plane and the binary
orbital plane. Note $S>B>0$ and the initial condition
$p_{0}=q_{0}=0$, thus the solution of $p$ and $q$ can be written as
\begin{eqnarray}
\label{solpq}
p&=&\frac{B\sin(2i_B)}{2f}\sin(f t),\nonumber\\
q&=&\frac{B\sin(2i_B)}{2B\cos2i_B+2S}\left[1-\cos(f t)\right],
\end{eqnarray}
where $f=\sqrt{(B\cos^2 i_B+S)(B\cos2 i_B+S)}$. The maximum value of
$i'$ (note that $i^{'}=\sqrt{p^{2}+q^{2}}$) is
\begin{equation}
\label{imax}
i'_{\rm{max}}=\frac{B\sin(2i_B)}{B\cos2 i_B+S}.
\end{equation}

As $S>>B$ shown in Figure \ref{tbs}, thus $f\sim S$ and
$i^{'}_{max}$ is as small as on an order of $o(B/S)$. It means that
the planetesimal will keep its orbital plane close to the disk
mid-plane, having the relative titled angle $i^{'}$ oscillating with
a frequency of $f\sim S$ and an amplitude of $\sim B/S$. Such an
analytical result is consistent with the hydrodynamical simulation
performed by \citet{Fra 11}, which has shown that the disk gravity
would try to pull the planetesimal orbit back to the disk mid-plane,
maintaining a small relative angle (see the figures 3 and 10 in
their paper).

Recalling the approximation (quantities that are $O(e^2)$, $O(i^2)$
or $O(ei)$ or higher are ignored) adopted before our derivation, we
thus emphasize that our analytical results about the evolution of
planetesimal inclination remain valid only if the planetesimal
eccentricity is not excited or remains at a low value. Such an
assumption, however, will break down if the Lidov-Kozai effect kicks
in. In the following subsection, we will address this issue,
deriving the conditions in which the Lidov-Kozai effect takes over
and planetesimal eccentricity is excited.

\subsection{Evolution of the Planetesimal Eccentricity}
Following \citet{Inn 97}, the Lagrange planetary equations
describing the evolution of the planetesimal's orbital eccentricity
($e$) and pericenter ($\omega$) can be written as
\begin{eqnarray}
\frac{de}{dt}&\sim&\frac{5B}{2}e\sin(2\omega)\sin^2i_{B} \label{dedt}\\
\frac{d\omega}{dt}&\sim&B\left(2-5\sin^2\omega\sin^2i_{B}\right)+D.\label{dwdt}
\end{eqnarray}
Compared to the equation (5) in the paper of \citet{Inn 97}, here we
add the term of contribution from the disk ($D$), ignore quantities
that are on an order of $o(e^{2})$ or higher because of the initial
circular planetesimal orbit, and take $i\sim i_{B}$ because $i^{'}$
is very small before $e$ is excited according to equation
(\ref{imax}). The disk contribution term ($D$) can be written as
(see the appendix for the detail of derivation)
\begin{eqnarray}
D\sim-T-\frac{SB\cos^2 i_B}{B\cos2 i_B+S}.
\label{D}
\end{eqnarray}
Note\footnote{Setting $i_{B}=0$ in equations (\ref{dwdt}) and
(\ref{D}), the binary and disk's contributions to
$\frac{d\omega}{dt}$ are $2B$ and $-B-T$ respectively, which are
obviously wrong, though their sum ($B-T$) is correct.}, as $i=0$ is
a singular point in the Lagrange planetary equation, thus equation
(\ref{dwdt}) and (\ref{D}) cannot be applied to the case of
$i_{B}=0$.

For the Lidov-Kozai effect to kick in, we expect $d\omega/dt \approx
0$. Using this condition to eliminate the variable $\omega$ in
equations (\ref{dwdt}) and (\ref{D}), we then have,
\begin{eqnarray}
\frac{de}{dt}&\sim& 5eB\sqrt{\left(\frac{2B+D}{5B}\right)
\left(sin^{2}i_{B}-\frac{2B+D}{5B}\right)}.
 \label{dedt2}
 \end{eqnarray}
In order to increase $e$, we need $de/dt>0$, which then leads to
\begin{equation}
\label{oo}
0<2+D/B<5\sin^2i_B.
\end{equation}
This is the condition for Lidov-Kozai effect to operate under the
gravity from both binary stars and the disk. For disk-free case,
i.e., $D=0$, then equation (\ref{oo}) is reduced to the classical
one, i.e., $i_{B}>\rm arcsin(\sqrt{2/5})\sim 39.2^{\circ}$.

As $D$ and $B$ are functions of the semimajor axis ($a$), inequation
(\ref{oo}) actually produces two critical semimajor axes,  a lower
limit of $a_{c1}$ and an upper limit of $a_{c2}$, which can be
derived from $2+D/B=0$ and $2+D/B=5\sin^{2}i_{B}$, respectively. If
the disk is not very tenuous, such as $f_{g}>0.1$, then $S\gg B$
holds and thus equation (\ref{D}) can be reduced to $D\sim
-T-B\cos^{2}i_{B}$.  In such a case, we can solve $a_{c1}$ and
$a_{c2}$ analytically if $k=-1$, \footnote{In order to analytically derive $a_{c1}$ and
$a_{c2}$ with moderate accuracy, we make the compromise that $B$ and $B_{1}$  has
the same dependency on a but a little difference in normalization, namely $B=(1+\eta) B_{1}$. Here in equation
(\ref{ac1}) $\eta=0.4$.   Note, equations (\ref{ac1}) and
(\ref{ac2}) cannot be applied to the disk free case by just setting
$f_{g}=0$ because we have presupposed that $f_{g}>0.1$. And for the case of $k\ne-1$, $a_{c1}$ and $a_{c2}$ should be solved numerically from Eqn.\ref{oo}.}
\begin{eqnarray}
a_{c1}&\sim&{4.17\rm{AU}}
\left[\frac{1}{f_g}\frac{M_B}{M_\odot}(\sin^2i_B+1)\right]^{-2/3}\left(\frac{a_B}{50\rm{AU}}\right)^{2},
\label{ac1}\\
a_{c2}&\sim& {4.17\rm{AU}}
\left[\frac{1}{f_g}\frac{M_B}{M_\odot}(1-4\sin^2i_B)\right]^{-2/3}\left(\frac{a_B}{50\rm{AU}}\right)^{2},\rm
if \, \, i_{B}<30^{\circ}, \nonumber  \\
& \sim& \rm \infty \rm \, \,  \, \ \ \ \ \ \ \ , if \, \, i_{B}\ge30^{\circ}. \label{ac2}
\end{eqnarray}

%

%
Comparing to the classical disk-free case, where Lidov-Kozai effect takes place only if $i_{B}>39.2^{\circ}$, here Lidov-Kozai effect (or eccentricity excitation) can occur for an arbitrary $i_{B}$, and the value of $i_{B}$ just determines the disk range ($a_{c1}<a<a_{c2}$) that subject to Lidov-Kozai effect.

\section{numerical study}

\begin{figure}
\begin{center}
\includegraphics[scale=0.75]{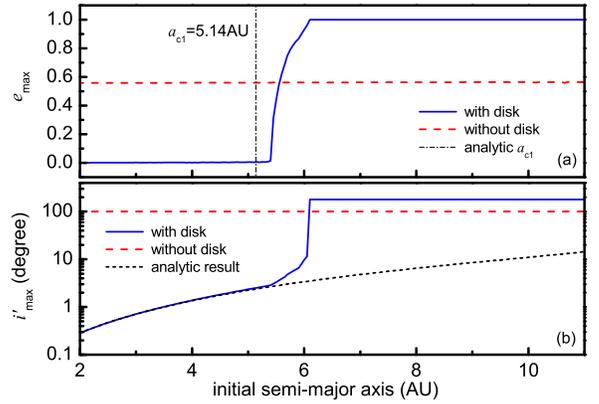}
\caption{Planetesimal's maximum eccentricities (panel (a)) and inclinations (panel (b)) as a function of its initial semi-major axis in the stander case. Red dashed
line indicates the results without including disk gravity. In panel (a), the vertical black
dash-dotted line indicates the analytical boundary of Kozai effect (Eqn.\ref{ac1}).
In panel (b), the black dashed line shows the analytical result from equation \ref{imax}.
}\label{sc1}
\end{center}
\end{figure}

\begin{figure*}
\begin{center}
\includegraphics[scale=0.80]{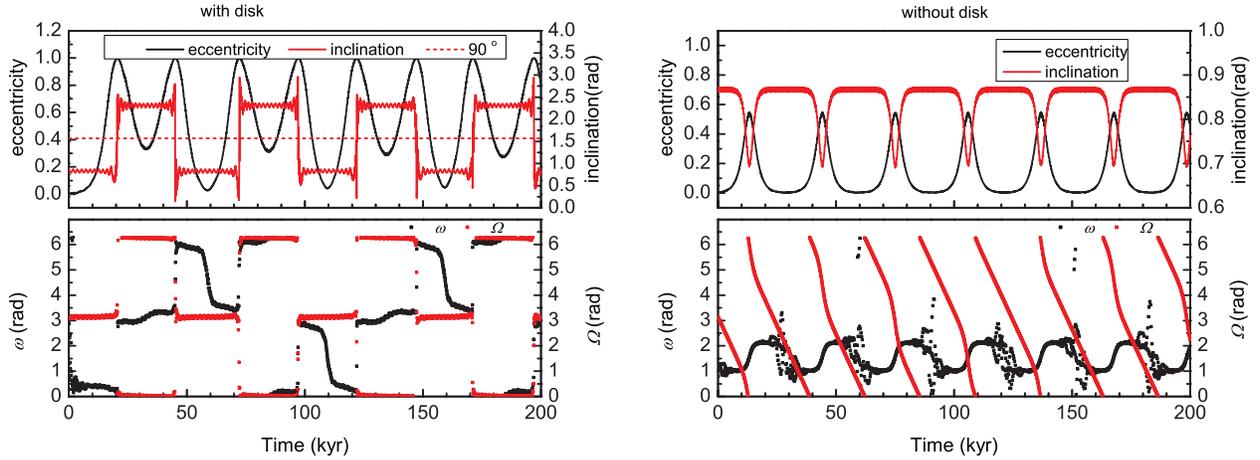}
\caption{Orbital evolution of a planetesimal with semimajor axis at
6.5 AU. All the orbital elements are in the binary coordinate. The
two left panels are results of the standard case, while the two
right panels are results for the case with the same binary
configuration but without including disk gravity. (Note the
different scales for the eccentricity and inclination scales in the
two plots.)}\label{evo}
\end{center}
\end{figure*}

In this section, we perform numerical simulations to test our
analytical results presented in section 3. Planetesimals are only
subject to the gravity from the binary stars and the
disk\footnote{In fact, planetesimals are also subject to the
hydrodynamical drag force from the gas disk. See section 5.1 for a
discussion of gas drag or see the paper of \citet{Xie 11} for a
detailed study of the effects of gas drag.}.  We calculate the
disk's gravity at lattice points in the $r^{'}$-$z^{'}$ plane before
orbital integrations and obtained the gravitational force at
arbitrary point by bicubic interpolation (see the appendix for a
detail description about computing disk gravity). The equations of
planetesimal motion are integrated using a fourth-order Hermit
method \citep{Kok 98}.

\begin{figure}
\begin{center}
\includegraphics[scale=0.75]{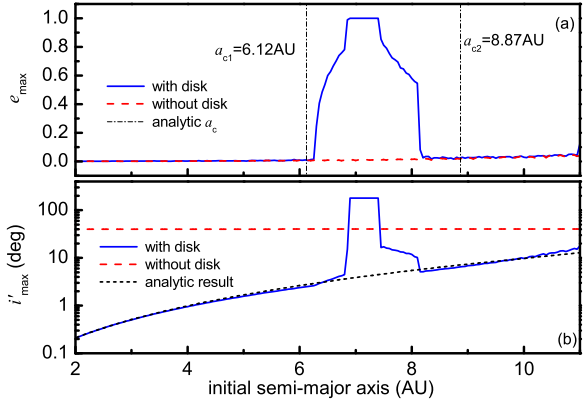}
\caption{Similar to Figure \ref{sc1} but with $i_{B}=20^{\circ}$}
\label{sc2}
\end{center}
\end{figure}

\subsection{Examples}

As a first example (hereafter referred to as the standard case), we
assume that $M_{B}=0.5 \rm M_\odot$,  $a_{B}=50$ AU, $i_{B}=50^{\circ}$ for the
binary and $f_{g}=1$, $k=-1$ for the disk. The results of this case is plotted in figure
\ref{sc1} and \ref{evo}.

In Figure \ref{sc1}, we plot the maximum orbital eccentricity
($e_{max}$) and inclination ($i^{'}_{max}$, in disk coordinates)
that the planetesimal achieved during its evolution as a function of
its orbital semimajor axis. As can be seen from Figure \ref{sc1}, in
the inner region, planetesimal eccentricities are not excited, and
they remain at very low inclinations with $i^{'}_{max}$ fitting well
with our analytical result (Eqn.\ref{imax}). In the outer region,
the Lidov-Kozai effect is switched on, and thus leads to large
planetesimal eccentricities ($e_{max}\sim1$) and inclinations
($i^{'}_{max}> 90^{\circ}$). The boundary that separates the inner
Kozai-off region and the outer Kozai-on region is roughly consistent
with the analytical estimate (Eqn.\ref{ac1}). In addition, we also
plot the results of the disk-free case as shown in the red dashed
line in Figure \ref{sc1}. Comparing the two cases of with and
without disk, we see that $e_{max}$ is much larger (close to unity)
in the former case.

In Figure \ref{evo}, for a specific planetesimal with semimajor axis
of 6.5 AU where the Lidov-Kozai effect should be switched on
according to Figure \ref{sc1}, we plot the temporal evolution of its
orbital eccentricity (e), inclination (i), longitude of periastron
($\omega$) and ascending node ($\Omega$) for the two cases with and
without disk. Note, here all the angular elements plotted in figure
\ref{evo} are  in the binary coordinate. In the case without disk,
the two right panels show the classical ``Lidov-Kozai'' cycle where
the eccentricity and inclination are evolving out of phase. However,
the situation is very different if the disk gravity is included in.
In such a case, as shown in the two left panels of Figure \ref{evo},
the planetesimal maintains its orbit around the initial one
($i=i_{B}, \Omega=180^{\circ}$) for a while at the beginning when
the eccentricity is not very high. As the planetesimal eccentricity
increases to the degree where $e \sim 1$, the planetesimal quickly
flips to a retrograde orbit but still in the same plane (the
mid-plane of gas disk) with $i\sim \pi-i_{B}$ and $\Omega=0 {\rm \,
or \, \pi}$. We note that such an orbital flip as well as the
associated high orbital eccentricity is very similar to the one
observed recently by \citet{Nao 11a, Nao 11b, LN 11}, where they
assume a non-zero eccentricity of the outer perturbing body, and the
orbital flip of the inner body is due to the so-called Eccentric
Lidov-Kozai Mechanism \citep{LN 11}. While in the present paper, we
assume a zero eccentricity of the outer perturbing body ($e_{B}=0$),
and thus the orbital flip observed in Figure \ref{evo} should be due
to the effect of disk gravity.

As a second example (hereafter referred to as the low inclination
case), we just change the binary orbital inclination to
$i_{B}=20^{\circ}$ and keep all the other parameters the same as in
the standard case. The result of this low inclination case is
plotted in figure \ref{sc2}. In contrast to the $i_{B} = 50^{\circ}$ case, here the
Lidov-Kozai effect can only take place within the region $a_{c1} < a
< a_{c2}$. This is consistent with our analytical results in
equations \ref{ac1} and \ref{ac2}.

\subsection{ Parameter exploration}
\begin{figure}
\begin{center}
\includegraphics[scale=0.75]{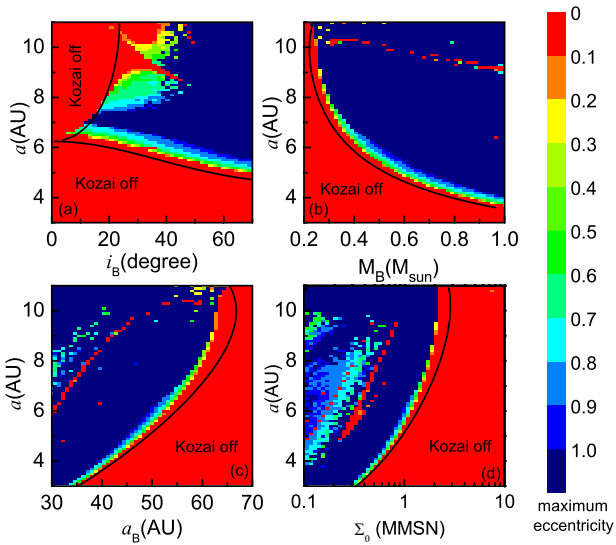}
\caption{Radial distribution of planetesimals' maximum eccentricity
and its dependency on $i_B$ (panel a), $M_B$ (panel b), $a_B$ (panel
c). and $f_g$ (panel d). Red-color regions mean the planetesimals
eccentricities $\sim$ 0, namely the Lidov-Kozai effect is switched
off. The black dashed lines indicate the analytical boundaries of
the Lidov-Kozai effect described by Equation \ref{oo} For all the four panels, $k=1$.} \label{grid}
\end{center}
\end{figure}

\begin{figure}
\begin{center}
\includegraphics[scale=0.75]{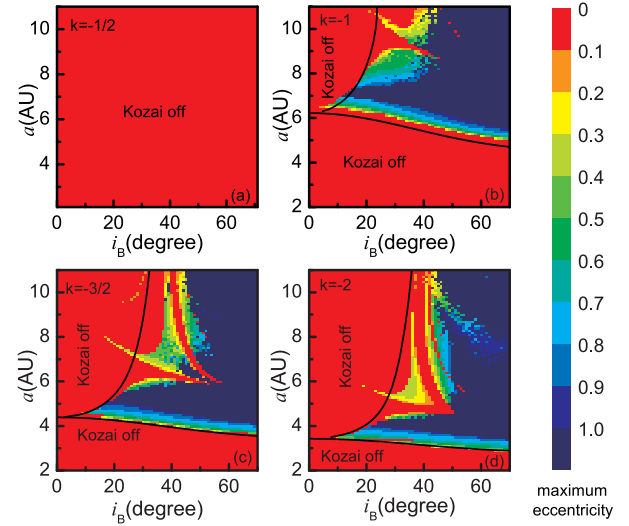}
\caption{Similar to the panel (a) of  Figure \ref{grid} but for cases of different $k$ values (the disk density slope). As can be seen, a flatter disk tends to be more efficient to suppress the Lidov-Kozai effect. This result is expected as flatting the disk profile is equivalent to increasing the density outside $r=1$ AU. } \label{grid2}
\end{center}
\end{figure}

In this subsection, we extend the standard case above by numerically
investigating the effects of other parameters, including $i_B$,
$a_B$, $M_B$, $f_g$, and $k$. We adopt the following strategy: To investigate
the effect of a given parameter, we set this parameter as the only
free one and fix all other parameters as the same as in the standard
case. The results are then plotted in Figure \ref{grid} and \ref{grid2}. The former shows
the radial distribution of planetesimal's maximum eccentricity and
its dependency on $i_{B}$ (top left), $a_{B}$ (bottom left), $M_{B}$
(top right) and $f_{g}$ (bottom right) in the case of $k=-1$. The latter just shows the dependency on $i_{B}$ but for the cases of different $k$ values. Some major features can be summarized as the following.

As shown in Figure \ref{grid},  (1) the Lidov-Kozai effect can be
switched on even with $i_B$ as small as $\sim 5^{\circ}$, although
the width of the Kozai- on region decreases as $i_B$ decreases. (2)
The Lidov-Kozai effect can be suppressed over a larger region if
either the mass of the companion star decreases, or the separation
of the companion and/or the density of the disk increases. (3) The
analytical results (dashed and solid lines, see also in
Eqn.\ref{ac1} and \ref{ac2}) approximate the numerical results in
the inner region with $a < 9 - 10$ AU. Beyond this, in the region
close to the disk outer boundary and the orbital stability boundary,
the deviation is large, indicating our analytical approximation is
not valid there. And (4) the boundaries that separate the Kozai-on
and Kozai-off regions are very steep; most planetesimal
eccentricities are either very high (close to 1) or very low (close
to 0) with planetesimals of moderate eccentricities being very rare.
The effect of the disk density slope $k$ can be seen from Figure \ref{grid2}, which shows (5) the Kozai-off region extends outwards more and more as $k$ increases, i.e., the disk radial density profile becomes more flat. In the case of $k=-1/2$, the Lidov-Kozai effect turns off in the whole disk.

\section{discussion}
\subsection{ Planetesimal Jumping and Pile-up (PJP) }
In the early stage, there must be a gas disk around the primary
star. The gas disk has crucial effects on the dynamics of
planetesimals through two factors. One is the gravity, which was
studied in detail in previous sections of this paper. The other one
is the hydrodynamic drag force, whose role has been investigated in
detail by \citet{Xie 11}.  In general, \citet{Xie 11} find that if
planetesimals are excited to orbits with very high inclinations
(relative to the disk plane) and eccentricities, they will be
subjected to very strong hydro-dynamic drag forces from the gas
disk, letting them jump inward and pile up, i.e., the so-called
Planetesimal Jumping and Pile-up (PJP) effect. Nevertheless, the
disk gravity is not included in by \citet{Xie 11}. In the following,
we show how the PJP effect is modified if both the gas drag and disk
gravity are included.

\begin{figure}
\begin{center}
\includegraphics[scale=0.75]{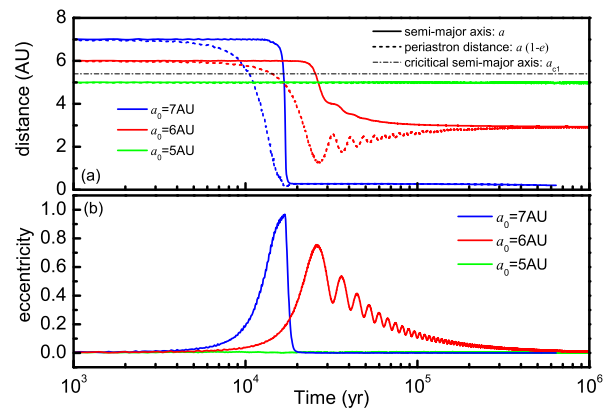}
\caption{Evolution of orbital semimajor axis  (also periastron, top
panel) and eccentricity (bottom panel) of three planetesimals with
initial semimajor axes of 5 (green) , 6 (red)  and  7 (blue) AU
respectively.  The binary configuration is set as in the standard
case described in section 4.1 but adding in the gas drag.  The black
dashed-dot line denotes the critical semimajor axis $a_{c1}=5.4$ AU
as measured from Figure \ref{sc1}.} \label{mig}
\end{center}
\end{figure}

\begin{figure}
\begin{center}
\includegraphics[scale=0.75]{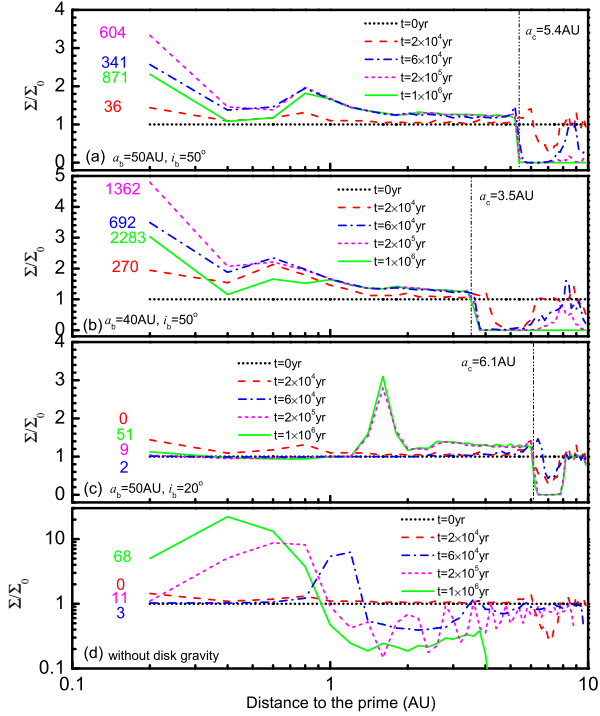}
\caption{Evolution of the local surface density enhancement
($\Sigma/\Sigma_{0}$) of the planetesimal disk in the standard case
(a), compact case (b), small inclination case (c) and the disk
gravity free case (d).  The vertical black dash-dotted lines show
the critical semi-major axis that separate the Kozai-on and
Kozai-off regions. The number on the left of each profile curve
denotes the total number of planetesimals which have migrated to the
innermost region within 0.2 AU.  Note the scale of the vertical axis
is different in the bottom panel as compared to those in the others.
}\label{pile1}
\end{center}
\end{figure}

We consider four cases, (a) the standard case as described in
section 4.1, (b) a more compact case$-$ similar to the standard case
but with $a_{B}$ decreasing to 40 AU, (c) a low inclination case$-$
similar to the standard case but with $i_{B}$ decreasing to
$20^{\circ}$, and (d) a disk gravity free case$-$ similar to the
standard case but the disk gravity is not included. In each case,
gas drag force is calculated by assuming a single planetesimal
radial size of 5 km and following the procedure as described in
section 2.2 of \citet{Xie 11}. The results are plotted in figures
\ref{mig} and \ref{pile1}.

%
%

In Figure \ref{mig},  we plot the evolution of orbital semimajor
axis (also periastron, top panel) and eccentricity (bottom panel) of
three planetesimals in case (a), i.e., the standard case. Three
planetesimals with near circular orbits starting from 5, 6  and 7
AU. Beyond $a_{c}=5.4$ AU, where the Lidov-Kozai effect is switched
on, the two planetesimals' eccentricities are excited and thus they
suffer significant gas drag force, leading to rapid inward
migration. The one starting from 7 AU is excited to extremely high
eccentricity and directly jumps to the innermost orbit, and the one
starting from 6 AU with modest eccentricity quickly migrates to 2-3
AU the central star. On the other side, for $a_c < 5.4$ AU, the
Lidov-Kozai effect is suppressed, and thus the planetesimal starting
from 5 AU does not suffer eccentricity excitation and hence does not
migrate.

In Figure \ref{pile1}, we plot the evolution of the local surface
density enhancement ($\Sigma/\Sigma_{0}$) of the planetesimal disk
for the four cases.  In each case, 5000 planetesimals treated as
test particle tracers, are initially distributed uniformly from 0.2
AU to 10 AU in the mid-plane of gas disk with circular orbits, thus
the initial profile follows a power law with a semi-major axis
dependence equal to $-1$. By tracing the radial distribution of
those planetesimals, we can calculate the local surface density
enhancement ($\Sigma/\Sigma_{0}$) of the planetesimal disk.  The
results are shown in Figure \ref{pile1} and can be summarized as
follows:

(1) As can be seen in the top panel (case a), planetesimals in the
outer Kozai-on region migrate into the inner Kozai-off region (in
fact, most are ``jumping'' as shown in Figure. \ref{mig} and pile up
there, leading to surface density enhancement in the innermost
region ($a < 0.2$ AU). This pile-up effect increases when the binary
separation, $a_B$, decreases (case b), because the outer Kozai-on
region is larger and thus more planetesimals can move in and pile
up. Conversely, when we reduce $i_B$ (case c), the pile-up effect is
reduced, and the pile-up region shifts outward to 1- 2 AU, because
the outer Kozai-on region shrinks and less particles are excited to
eccentricity close to 1 (see Figure. \ref{sc2}).

(2) If the disk gravity is not included (case d), then the situation
reduces to the situation considered in \citet{Xie 11}. In such a
case, the Lidov-Kozai effect can only be suppressed by the gas
damping and this only in the very inner region within 1-2 AU, where
gas density is high. Beyond 1-2 AU, planetesimals experience the
Lidov-Kozai effect and most (if not all) will migrate inward and
pile up within $\sim 0.2 - 1.0$ AU, leading to an average local
density enhancement of $\Sigma/\Sigma_0 \sim 10$ (see also in the
Figure. 9 of \citet{Xie 11}). However, we note that there are many
fewer planetesimals piling up within region $< 0.2$ AU in case (d)
than in case (a). The reason is that the planetesimal eccentricity
(see Figure. \ref{evo}) in case (d) is not high enough to let
planetesimals directly jump into the innermost region $< 0.2$ AU.

In a word, the role of the disk gravity playing in the PJP effect
can be summarized as the following. On one hand, disk gravity
reduces the average PJP effect because it reduces the Kozai-on
region in the outer disk. However, on the other hand, the disk
gravity significantly enhances the PJP effect in the innermost
region ($< 0.2$ AU) as it increases the orbital eccentricities of
planetesimals in the Kozai-on region to values close to 1.

\subsection{Effects of Planetesimal Collisions}
In this paper, the planetesimals are treated as test particles and their mutual collisions are ignored. As planetesimals jump inward, their orbital eccentricities are very high and thus they are potentially subject to collisions of very high relative velocities, which can entirely disrupt themselves. To know how relevant the collisions could be, we estimate the collisional timescale ($t_{\rm col}$) first. Following \citet{Xie 10b}, $t_{\rm col}$ in an inclined binary system can be estimated as \footnote{Combine Eqn (3), (6) and (7) in \citet{Xie 10b}}.
\begin{eqnarray}
t_{\rm col}\sim{4\over3}\times10^4 f_{\rm g}^{-1} f_{\rm ice}^{-1}({i_{\rm B}\over \rm 1^{\circ}})({a\over \rm AU})^3({M_{\rm A}\over \rm M_\odot})^{-1/2}({R_{\rm p}\over \rm km})  \ \ \ \rm yr,
\end{eqnarray}
where $f_{\rm ice}$ is solid density enhancement beyond the ice line, $R_{\rm p}$ is the planetesimal radii. Taking typical parameters, i.e., $f_{\rm g}=1.0$,$f_{\rm ice}=4.2$, $i_{B}=50^{\circ}$, $M_{\rm A}=M_{\odot}$, $R_{\rm p}=5$ km, it gives $t_{\rm col}\sim8\times10^{5} (a/\rm AU)^{3}$ yr. As planetesimals complete their jumps typically in a timescale of $10^{4}-10^{5}$ yr shown in Figure \ref{mig} and \ref{pile1}, thus we conclude that collisions have little effects before or under the process of planetesimal jumping, but they do play important roles after plantesimals jumping inside 1 AU. Actually, this expectation is confirmed by the simulations in \citet{Xie 11}. As shown in the figure 13 of their paper, in the first $10$ yr, the collisional velocity is very high but the collisional frequency is rather low. Afterwards, collisions become more frequent as more planetesimals pile up in the inner region. At the same time, planetesimals are damped to near coplanar and circular orbits, leading to a friendly condition for subsequent planetesimal growth by mutual collisions.

\subsection{Implication to the Formation of Hot Jupiters}
The Lidov-Kozai effect induced by a companion star in a binary
system has been suggested as an important mechanism for the
formation of hot Jupiters \citep{WM 03,  FT 07}. If a planet's
eccentricity is high enough that its periastron is very close to the
star (say $< 0.1$ AU) during the Kozai cycle, then tidal dissipation
can kick in, which may circularize and shrink the planet's orbit,
finally letting it become a hot planet. However, to induce such a
high eccentricity by the classical Lidov-Kozai effect, it needs an
extremely misaligned configuration{\footnote{The critical $i_{B}$
can be lower if one considers the effect of the binary eccentricity
\citep{LN 11} }} (say $i_{B}>85^{\circ}$, according to \citet{WM
03}), which is not common and thus lowers the chance of forming a
hot-Jupiter. As estimated by \citet{Wu 07}, such a ``stellar Kozai''
mechanism can only produce 10\% hot Jupiters.

Nevertheless, the situation will be different if the disk gravity is
included in the Lidov-Kozai effect. In such an case, almost in the
whole Kozai-on region, the eccentricity can be excited to be an
arbitrarily high value even with very low initial binary inclination
($i_{B}$), which produces many more ``hot planetsimals'' ($a<0.2$
AU) as shown in Figure \ref{pile1}.  Similarly, if our model and
results can be applied to a Jupiter-like planet \footnote{Note, the
results might be different because the giant planet can
significantly affect the gas disk, e.g., opening a gap.} (by
replacing the gas drag damping with tidal damping in figure
\ref{pile1}) , it should also produce many more hot Jupiters.  The
key issue is,  to what degree the production rate of hot jupiter can
increase via the above ``modified stellar Kozai'' mechanism.  We
will address this in detail in a forthcoming paper.

\begin{figure}
\begin{center}
\includegraphics[scale=0.8]{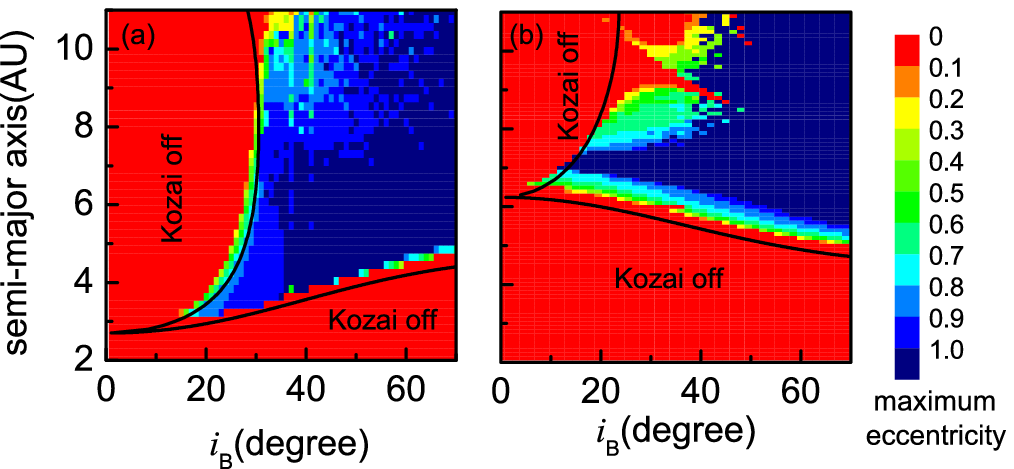}
\caption{Panel (a). Maximum eccentricities distribution in the plane
of $i_B-a$ if the precession of the disk is considered. The black
solid lines indicate the analytic boundaries of Kozai-on region.
Panel (b). Same as the top left panel of Fig 5. Maximum
eccentricities distribution if the precession of the disk is not
considered.}\label{pre3}
\end{center}
\end{figure}

\begin{figure}
\begin{center}
\includegraphics[scale=0.95]{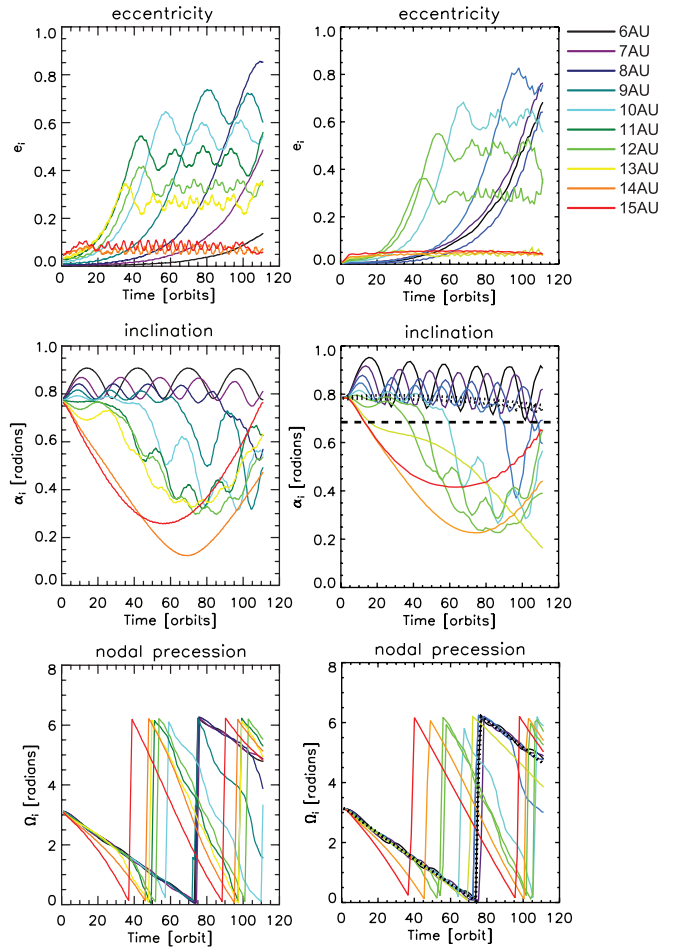}
\caption{Comparison between our results (left three panels) and
those in  figure 10 of \citet{Fra 11} (right three panels). From top
to bottom, they are evolutions of planetesimals' eccentricities,
orbital inclinations (relative to the binary orbital plane), and
nodal precession, respectively. The line color indicate the
planetesimal's initial semi-major axis. The initial setup, including
the configuration of the binary and disk is adopted from the model 3
of \citet{Fra 11} (see section 3.2 and table 2 of their
paper).}\label{comp}
\end{center}
\end{figure}

\subsection{Disk Precession}
In this paper, we assume that the gas disk is non-evolving and
axisymmetric, which is apparently a crude approximation. In fact,
the gas disk (if it is not entirely disrupted) should undergo a near
rigid body precession \citep{Lar 96, FN 10}, and the
precession rate can be estimated as
\begin{equation}
\dot{\Omega}_{d}=-\left(\frac{3GM_B}{4a_B^3}\cos i_B \frac{\int\Sigma_{g} r^3
dr}{\int\Sigma_{g} \Omega_{\rm k, mid} r^3 dr}\right).\label{omgdot}
\end{equation}
For the standard case considered in this paper, equation
\ref{omgdot} gives $\dot{\Omega_{d}}\sim -2.6\times10^{-4}{\rm{rad\,
yr}}^{-1}$. Adding such a rigid precession to the gas disk, we
re-run the simulations shown in the top-left panel of Figure
\ref{grid} and plot the results in Figure \ref{pre3}. The two black
solid curves in Figure \ref{pre3} are two critical semimajor axes
($a_{c1}$ and $a_{c2}$) derived from equation \ref{oo} (not from
Eqn.\ref{ac1} and \ref{ac2}) by assuming
\begin{equation}
D\approx T+B\cos^2 i_B+\cos i_B\dot{\Omega}_{d}.
\label{dd}
\end{equation}
Although equation \ref{dd} is a very crude approximation, it
produces reasonable $a_{c1}$ and $a_{c2}$ which fit the numerical
results as well as shown in Figure \ref{pre3}. Furthermore, both the cases with and without disk precession (comparing the two panels of Fig.\ref{pre3})
produce some similar features, such as: (i) in
the central regions of the disk, the Lidov-Kozai effect can be
switched on at very low inclinations, and (ii) once the Lidov-Kozai
effect is switched on, the planetesimal eccentricities can be much
higher (most are close to 1) than those in the case without disk
gravity.

\subsection{Comparison to the Hydrodynamical Results}
In order to further examine the validity of our disk model, we
compare the results of our model to the hydrodynamical results given
by \citet{Fra 11}. We adopt the same initial set up as in the
simulations shown in the figure 10 of \citet{Fra 11} and run the
simulation with our model and numerical method described in Appendix
D. The comparison results are plotted in Figure \ref{comp}. As can
be seen, the results computed by our model are generally consistent
with the hydrodynamical results of \citet{Fra 11}. Given such a
comparison, we then feel confident of the results shown in other
places of this paper.

\section{summary}
In this paper, we investigated the effects of gas-disk gravity on
planetesimal dynamics in inclined binary systems using both analytic
and numerical methods. Our major conclusions are summarized as the
following.

Analytically, we derive that the planetesimal inclination follows a
small amplitude oscillation around the mid-plane of disk (see
Eqn.\ref{solpq} and \ref{imax}) if the Lidov-Kozai effect is
suppressed and thus planetesimal eccentricity is not excited.
Furthermore, we derive the threshold condition (see Eqn.\ref{oo},
\ref{ac1} and \ref{ac2}) in which the Lidov-Kozai effect switches
on. We find the Lidov-Kozai effect can operate at very low
inclinations if the disk gravity is considered, although the radial
extent of the Kozai-on region is much smaller.

Numerically, we confirm our analytical results over a very large
parameter space by considering the variation of $i_{B}$, $a_{B}$,
$M_{B}$, $f_{g}$. We find that the disk gravity narrows down the
Kozai-on region, but at the same time significantly increases the
maximum eccentricity (close to 1) of planetesimals in the Kozai-on
region (see Figure \ref{sc1}). Such high planetesimal eccentricities
usually accompany orbital flipping (see Figure \ref{evo}), i.e.,
planetesimal orbits flip  back and forth between prograde to
retrograde.

Applying the effects of disk gravity to the planetesimal
jumping-piling (PJP) process. We find that, on the average over the
disk, disk gravity reduces the PJP effect. However, PJP effect is
significantly enhanced in the innermost region within 0.2 AU (see
Figure \ref{pile1}) . In addition, given the extremely high
eccentricity under the effects of disk gravity, we believe that the
production rate of hot-Jupiters via the ``stellar Kozai'' mechanism
could be increased.

\acknowledgments

We are grateful to Dr. Matthew Payne, Dr. Sverre Aarseth and Dr.
Yanqin Wu for useful discussions and suggestions. This work is
supported by the National natural Science Foundation of China
(Nos.10833001, 10778603, and 10925313), and the National Basic
Research Program of China(No.2007CB814800).

\appendix
\section{A.\, The disturbing function of the disk}
According to Nagasawa et al. 2000, taken to second order in $e$ and
$i'$, the disturbing function caused by the disk can be expressed as
\begin{equation}\label{rd}
R_{D}=-\frac{na^2}{2}\left[T(a)e^2+S(a)i'^2\right],
\end{equation}
$T(a)$ and $S(a)$ are given using an integral of cylindrical
coordinates $(r',\phi',z')$:
\begin{eqnarray}\label{tands}
T(a)&=&\frac{1}{2n}\int_{r_{\rm{in}}}^{r_{\rm{out}}}\int_{-\infty}^{\infty}\int_{0}^{2\pi}\left[\frac{3-2r'\cos\phi'/a}{\Delta^3}\right.\\\nonumber
&&-\left.\frac{3(a-r'\cos
\phi')^2}{\Delta^5}\right]G\rho_g(r',z')r'dr'd\phi'dz',\nonumber
\\\nonumber
S(a)&=&\frac{1}{2n}\int_{r_{\rm{in}}}^{r_{\rm{out}}}\int_{-\infty}^{\infty}\int_{0}^{2\pi}\left(\frac{r'\cos\phi'/a}{\Delta^3}-\frac{3z'^2}{\Delta^5}\right)G\rho_g(r',z')
r'dr'd\phi'dz',
\end{eqnarray}
where $\Delta=(a^2+r'^2+z'^2-2ar'\cos\phi')^{1/2}$\\

\section{B.\, Inclination evolution equation}

Ignoring $e^2$ and higher order terms in the disturbing function
$R$, perturbation function relating to the inclination has the form
\begin{equation}\label{rishort}
R=\frac{na^2}{2}\left[-S(a)i'^2-B(a)\sin^2i\right],
\end{equation}
where $i$ is the inclination in the binary coordinate and $i'$ is
that in the disk coordinate. In the binary coordinate system, the
xy-plane is the binary¡¯s orbital plane, and the x-axis is the
ascending node of the companion with respect to the disk. In the
coordinate system of the disk, the x-axis is same as that of the
binary coordinate system, and the xy-plane is the mid-plane of the
disk.

According to the geometrical relationship between the two
coordinate, we have
\begin{equation}
\begin{pmatrix} \sin i\sin\Omega\\ -\sin i\cos\Omega \\ \cos i\end{pmatrix}
=
\begin{pmatrix} &1&0&&0\\
&0&\cos i_B&&\sin i_B\\
&0&-\sin i_B&&\cos i_B \end{pmatrix}\begin{pmatrix} \sin i'\sin\Omega'\\
-\sin i'\cos\Omega' \\ \cos i'\end{pmatrix}.
\end{equation}
It is easy to obtain
\begin{equation}
\left\{\begin{aligned}&\sin i\sin\Omega&=&\sin i'\sin\Omega',\\
&\sin i\cos\Omega&=&\sin i'\cos\Omega'\cos i_B-\cos i'\sin i_B,\\
&\cos i&=&\sin i'\cos\Omega'\sin i_B+\cos i'\cos i_B.
\end{aligned}\right.
\end{equation}
Then
\begin{eqnarray}
\sin^2i&=&1-\sin^2i'\cos^2\Omega'\sin^2 i_B-\cos^2 i'\cos^2
i_B-2\sin i'\cos\Omega'\sin i_B\cos i'\cos i_B \nonumber \\
&=&\sin^2i_B+\sin^2i'\cos^2i_B-\sin^2i'\cos^2\Omega'\sin^2 i_B-\sin
i'\cos i'\cos\Omega'\sin2i_B\nonumber \\
&=&\sin^2i_B+\sin^2i'\sin^2\Omega'\cos^2i_B+\sin^2i'\cos^2\Omega'\cos
2i_B-\sin i'\cos i'\cos\Omega'\sin2i_B \nonumber \\
&=&i'^2\sin^2\Omega'\cos^2i_B+i'^2\cos^2\Omega'\cos
2i_B-i'\cos\Omega'\sin 2i_B+\sin^2i_B+o(i'^3).
\end{eqnarray}
If we ignore $i^3$ and higher order terms, the relationship becomes
\begin{equation}\sin^2i=p^2\cos^2i_B+q^2\cos2i_B-q\sin(2i_B)+\sin^2i_B,\end{equation} where
$p=i'\sin \Omega'$ and $q=i' \cos \Omega'$. Thus the perturbation
function becomes
\begin{equation}\label{ri2}
R=\frac{na^2}{2}\left[-(B\cos^2i_B+S)p^2-(S+B\cos2i_B)q^2+B\sin(2i_B)q-B\sin^2i_B\right]
\end{equation}
Using Lagrange's equations of motion, the evolution of the
inclination is given by
\begin{eqnarray}\label{dpdq}\frac{dp}{dt}&=&-\left(B\cos2 i_B+S\right)q+\frac{B}{2}\sin(2i_B),\nonumber\\
\frac{dq}{dt}&=&\left(B\cos^2i_B+S\right)p.
\end{eqnarray}
\\

\section{C.\, THE CONTRIBUTION OF DISK TO THE PERIASTRON
PRECESSION} The disturbing function of the disk has the form
\begin{equation}\label{rd2}
R_{D}=-\frac{na^2}{2}\left[T(a)e^2+S(a)i'^2\right].
\end{equation}
Using Lagrange's equations of motion, and ignoring the $e^2$ term,
we can the expression for the evolution of $\omega$ cause by the
disk
\begin{equation}
\left(\frac{d\omega}{dt}\right)_{disk}=-T+\frac{1}{2}S\cot
i(\partial i'^2/\partial i)\equiv D.
\end{equation}
where $i$ and $i'$ are the inclinations in the companion coordinate
system and disk coordinate system. Proceeding with the same method
as used in Appendix A,  we have
\begin{equation}
\sin^2i'=\sin^2 i\sin^2 \Omega\cos^2i_B+\sin^2i \cos^2
\Omega\cos2i_B +\sin i\cos i \cos \Omega\sin 2i_B+\sin^2i_B,
\end{equation}
then we can obtain that
\begin{eqnarray}
\partial (\sin^2i')/\partial i&=&\sin 2i\sin^2 \Omega\cos^2i_B+\sin 2i \cos^2
\Omega\cos2i_B +\cos 2i \cos \Omega\sin 2i_B\nonumber \\
&=&2\cot i\left(\sin^2i\sin^2 \Omega\cos^2i_B+\sin^2i \cos^2 \Omega\cos2i_B
+\sin i\cos i \cos \Omega\sin 2i_B+\sin^2i_b\right)-\cos \Omega\sin
2i_B-2\cot i\sin^2i_b\nonumber \\ &=&\left[2\sin^2i'\cos i-2\sin
i_B\left(\sin i\cos \Omega\cos i_B+\cos i\sin i_B\right)\right]/\sin
i\nonumber \\ &=&\left(2\sin^2i'\cos i-2\sin i'\cos \Omega'\sin
i_B\right)/\sin i.
\end{eqnarray}
Because initially $i'=0$,we ignore $o(i'^2)$ term and have
\begin{equation}
\partial (i'^2)/\partial i=-2i'\cos\Omega'=-2q.
\end{equation}
We have obtained previously that
\begin{equation}q=i'\cos\Omega'=\frac{B\sin(2i_B)}{2B\cos2i_B+2S}\left[1-\cos(f t)\right].\end{equation}
For the case $S>>B$, the timescale of the evolution of $i'$ and
$\Omega'$ is much shorter than the Kozai timescale. Thus we replace
$q$ with its average value
\begin{equation}<q>=\frac{B\sin(2i_B)}{2B\cos2i_B+2S}=i'_{max}/2.\end{equation}
and $D$ becomes
\begin{equation}D=-T-\frac{1}{2}Si_{max}\cot i=-T-\frac{SB\cos^2 i_B}{(B\cos2
i_B+S)}.\end{equation}
\\

\section{D. \, Gravitational force of the disk}
According to Nagasawa et al. (2000), the potential of the disk at
$(r,\phi,z)$ is
\begin{eqnarray}
\label{potential}
V=G\int_{r_{\rm{in}}}^{r_{\rm{out}}}\int_{-\infty}^{\infty}\int_{0}^{2\pi}\frac{\rho(r',z')r'd\phi'dz'dr'}{(r^2+r'^2-2rr'\cos\phi'+(z-z')^2+\epsilon)^{1/2}},
\end{eqnarray}
where $r_{\rm{in}}$, and $r_{\rm{out}}$ are the inner edge and the
outer edge of the disk,  respectively, and $\epsilon$ is a softening
parameter used to avoid a singularity. For reasons of efficiency and
precision, we set it to be $1\times10^{-7}$. Derivative of the
potential with respect to $r$ or $z$ yields the r or z component of
the disk¡¯s gravity,
\begin{eqnarray}
\label{force}
F_r=G\int_{r_{\rm{in}}}^{r_{\rm{out}}}\int_{-\infty}^{\infty}\int_{0}^{2\pi}\frac{\rho(r',z')(r-r'\cos\phi')r'd\phi'dz'dr'}{(r^2+r'^2-2rr'\cos\phi'+(z-z')^2+\epsilon)^{3/2}},\nonumber\\
F_z=G\int_{r_{\rm{in}}}^{r_{\rm{out}}}\int_{-\infty}^{\infty}\int_{0}^{2\pi}\frac{\rho(r',z')(z-z')r'd\phi'dz'dr'}{(r^2+r'^2-2rr'\cos\phi'+(z-z')^2+\epsilon)^{3/2}},
\end{eqnarray}
We numerically integrated equation (\ref{force}) using closed
Newton-Cotes formulas with Bode¡¯s rule \citep{Pre 92}.  Since the
integration costs too much CPU time, we can not do it for each
orbital integration step. Instead, we calculated the disk¡¯s gravity
at lattice points in the r-z plane before starting the orbital
integrations and obtained the gravitational force at arbitrary
points during the orbital integration by performing bicubic
interpolations \citep{Pre 92} using the value at lattice points.

\end{document}